\documentstyle[prl,twocolumn,aps]{revtex}
 
\begin{document}

\draft

\title{Quantum error correction in the presence of spontaneous emission.}
\author{M. B. Plenio, V. Vedral, and P. L. Knight}
\address{Blackett Laboratory, Imperial College London, 
London SW7 2BZ, England}
\date{\today}
\maketitle
\begin{abstract} 
We present a quantum error correcting code that is invariant under
the conditional time evolution between spontaneous emissions and which
can correct for one general error. The code presented here 
generalizes previous error correction codes in that not all errors 
lead to different error syndromes. This idea may lead to shorter 
codes than previously expected.
\end{abstract}
\section{Introduction}
With the discovery of an algorithm to factorize a large number on a 
quantum computer in polynomial time instead of exponential time as 
required by a classical computer \cite{Shor1}, the question of how to 
implement such a quantum computer has received considerable attention
\cite{Cirac1}. However, realistic estimates soon showed that decoherence
processes and spontaneous emission severely limit the bit size of the 
number that can be factorized \cite{Plenio1,Plenio2}. It has become clear that 
the solution to the problem does not lie in an increase in the lifetime 
of the transitions used in the computation. Attention has now shifted 
towards the investigation of methods to encode qubits such that the 
correction of errors due to interaction with the environment becomes 
possible. In a number of recent publications, possible encoding schemes 
have been considered and theoretical work has been undertaken to elucidate
the structure of quantum error correction codes
\cite{Shor2,Chuang1,Calderbank1,Steane1,Steane2,Laflamme1,Ekert1,Vaidman1,
Knill1,Bennett1,Barenco1,Calderbank2,Shor3,Steane3,DiVincenzo1,Gottesmann2,
Gottesmann1,Cleve1}. 
However, we show that these codes do 
not perfectly correct errors due to the conditional time evolution 
\cite{Dalibard1} between spontaneous emissions . This has the effect that 
for example the encoded lower state of a qubit, which, if unencoded, is
not influenced by the conditional time evolution, acquires an error due to 
the conditional time evolution. We then proceed to construct a code that 
is able to correct {\em one} general error and is able to correct to 
{\em all orders} the errors due to the conditional time evolution between 
spontaneous emissions. By one general error we mean an arbitrary one 
bit operation acting on a single bit of the code. The conditional time
evolution, however, contains terms that act on many qubits. Our code 
proposed in this paper is the first code that has the ability to correct 
a special kind of error (here due to the conditional time evolution) to 
all orders. This is an interesting feature, as one would be interested to 
correct those errors which frequently occur to higher order than rare errors.
The code presented here is optimal in the sense that it uses the smallest
possible number of qubits required to perform its task (correcting one 
general error and all errors due to the conditional time evolution). 
\section{Single error correcting codes}
Several codes have been proposed to encode one qubit which can correct one 
general error, i.e. amplitude and phase error or a
combination of both applied to the same qubit. An example \cite{Laflamme1}
of such a code is one where state $|0\rangle$ is represented by
\begin{eqnarray}
	|0_L\rangle &=& |00000\rangle + |11100\rangle - |10011\rangle
	  		            -|01111\rangle
	\nonumber\\
 		    && + |11010\rangle + |00110\rangle + |01001\rangle
				    +|10101\rangle
	\label{1a}
\end{eqnarray}
and the state $|1\rangle$ by
\begin{eqnarray}
	|1_L\rangle &=& |11111\rangle - |00011\rangle + |01100\rangle
				      - |10000\rangle
	\nonumber\\
		    && - |00101\rangle + |11001\rangle + |10110\rangle
				      - |01010\rangle \;\; ,
	\label{1b}
\end{eqnarray}
where the subscript $L$ indicates that the encoded state $|i_L\rangle$ 
differs from the initial state $|i\rangle$. We omit the obvious normalization
factor in the states $|0_L\rangle$ and $|1_L\rangle$ throughout this
letter as they are irrelevant for the present analysis. Starting with a 
state $|\psi\rangle = \alpha|0\rangle + \beta|1\rangle$, this is encoded 
as $|\psi_L\rangle = \alpha|0_L\rangle + \beta|1_L\rangle$. If the state 
suffers an amplitude error $A_i$ (which acts as a NOT operation on 
qubit $i$) or a phase error $P_i$ (which gives the upper state of qubit $i$ 
a minus sign) or the combination $A_iP_i$ of both to the $i$th qubit of 
$|\psi_L\rangle$ it is possible to reconstruct the initial state 
$|\psi\rangle$. The code given in eqs. 
(\ref{1a}) - (\ref{1b}) has the attractive feature that it is optimal 
in the sense that it only requires $5$ qubits which can be shown to be 
the minimal possible number \cite{ Knill1}. Using ideas similar
to classical error correcting codes one can estimate that if one wants 
to encode $l$ qubits in terms of $n$ qubits in such a way that one can
reconstruct the state after $t$ general errors, then the inequality 
\begin{equation}
	2^l \sum_{i=0}^{t} 3^i \left(n \atop i \right) \le 2^n
	\label{3}
\end{equation} 
has to be satisfied \cite{Ekert1}. The bound eq. (\ref{3}) is related
to the sphere packing bound in classical coding theory \cite{Pless1}.
The reason for that is that eq. (\ref{3}) was obtained using 
the assumption that different errors lead to different mutually 
orthogonal error syndromes. However, we will later see that the code 
presented in this paper (like the one presented in \cite{Shor2}) in 
fact violates this assumption which shows that it may be possible to find 
codes that go beyond eq. (\ref{3}).

The code given in eqs. (\ref{1a}) - (\ref{1b}) does not correct
for multiple errors. Especially, it is not able to correct to all orders
for errors that arise due to the conditional time evolution between 
spontaneous emissions. The conditional time evolution between 
spontaneous emissions is unavoidable and it differs from the unit
operation because the fact that no spontaneous 
emission has taken place provides information about the state of the
system and therefore changes its wave function. The conditional time 
evolution of the system under the assumption that no spontaneous emission 
has taken place is given by the non-unitary time evolution operator
$exp\{-i\,H_{eff} t/\hbar\}$ \cite{Dalibard1}. For the case that the 
qubits are not driven by external fields we obtain for the code given 
in eqs. (\ref{1a}) - (\ref{1b}) the effective Hamilton operator
\begin{equation}
	H_{eff} = \sum_{i=1}^{5} -i\,\hbar\Gamma\sigma_{11}^{(i)} \;\; ,
	\label{4}
\end{equation}
where $\sigma_{11}^{(i)}$ is the projector $|1\rangle\langle 1|$ onto the 
excited state of the $i$th qubit leaving all other qubits unaffected. 
$2\Gamma$ is the Einstein
coefficient of the upper level $1$ of the qubits. If we apply the 
conditional time evolution $exp(-i\,H_{eff} t/\hbar)$ to the encoded
state
\begin{equation}
	|\psi_L\rangle = \alpha|0_L\rangle + \beta|1_L\rangle
	\label{5}
\end{equation}
and subsequently apply the appropriate error correction procedure 
for this $5$-bit code \cite{Laflamme1} we do {\em not} recover the 
original state. This becomes obvious in the special case 
$\Gamma t \gg 1$ in which one obtains
\begin{eqnarray}
	|\psi_C\rangle &=& |00000\rangle + |00010\rangle + |01000\rangle
			   - |01110\rangle
	\nonumber\\
	               &&  + |10000\rangle + |10010\rangle + |11000\rangle
			 + |11110\rangle \;\; .
	\label{6}
\end{eqnarray}
This shows that this $5$-bit code is not able to correct errors due to the 
conditional time evolution exactly. Especially striking is the effect
when we assume that $\beta=0$, ie. we encode the (stable) ground state.
The conditional time evolution then leads to no errors in the unencoded 
state while it changes the encoded state such that it cannot be corrected 
perfectly anymore. Note, however, that the error introduced by the conditional 
time evolution is, for short times, of fourth order. If, however, a 
spontaneous emission (or any other kind of error) occurs then a subsequent 
conditional time evolution induces contributions which after error
correction lead to second order errors in the state. Our
code presented later in this paper preserves the encoded state in both 
cases perfectly, ie. to all orders.

The reason that the code eqs. (\ref{1a})-(\ref{1b}) cannot perfectly 
correct errors
due to the conditional time evolution derives from the fact that the 
words (product states) of which the code consists do not all have the same 
number of excited states. This leads to a difference in the rate at which 
the amplitude of these states decays. The amplitude of $|00000\rangle$
remains unchanged under the conditional time evolution while the amplitude
of $|11100\rangle$ for example decreases at a rate $exp(-3\Gamma t)$. This 
can be seen as a multiple amplitude error with which the code can 
not cope. This problem is not restricted to the $5$-bit code given in 
\cite{Laflamme1} but is present in all other previously proposed codes. 
It should be noted that it is not necessary to observe the system for 
these conclusions to hold. If we do not observe the system, it then has 
to be described by a density operator, whose time evolution follows the 
appropriate Bloch equations. This time evolution can in principle be 
decomposed into individual trajectories each of which consists of no--jump 
evolutions interrupted by spontaneous emissions \cite{Dalibard1}. For 
each of these trajectories our considerations above hold and therefore 
also hold for the incoherent sum of these trajectories which make up the
ensemble. Therefore our error correction code is not restricted to a 
particular 
measurement scheme such as for example the detection and reconstruction scheme 
discussed in \cite{Mabuchi1}, where it is necessary to detect individual 
quantum jumps. Nevertheless such a detection of individual jumps 
would improve the performance of our code, as that would exclude the 
contribution of multiple quantum jumps with which our code cannot cope. 
This would enhance the importance of the conditional time evolution as a error 
source compared to other sources and it is here where our code is superior to 
previous codes.
\section{Correcting spontaneous emission}
The discussion of the last section shows that it is of some interest to 
construct a quantum error correcting code that corrects errors due to the 
conditional time evolution to all orders. This is possible, and in the 
following we present such a quantum error correcting code.

The following code was constructed starting from the code 
(\ref{1a})-(\ref{1b}). State $|0\rangle$ is encoded as
\begin{eqnarray}
	|0_L\rangle &=& |00001111\rangle + |11101000\rangle - 
			|10010110\rangle - |01110001\rangle 
	\nonumber\\
    \hspace*{-.1cm}+&& \hspace*{-.5cm}
		        |11010100\rangle + |00110011\rangle + 
                        |01001101\rangle + |10101010\rangle \;\; ,
	\label{8}
\end{eqnarray}
while state $|1\rangle$ is encoded as
\begin{eqnarray}
	|1_L\rangle &=& |11110000\rangle - |00010111\rangle + 
		        |01101001\rangle -|10001110\rangle
	\nonumber\\
    \hspace*{-.1cm}- && \hspace*{-.55cm} 
	                |00101011\rangle + |11001100\rangle + 
		        |10110010\rangle - |01010101\rangle
	\; .
	\label{9}
\end{eqnarray}
The state eq. (\ref{8}) encoding the logical $0$ was obtained in the 
following way. We started with state eq. (\ref{1a}) and for each word, e.g.
$|11100\rangle$ we constructed the bitwise inverse, i.e. $|00011\rangle$.
We concatenated the two words where the second one is taken in reverse bit
order to obtain $|1110011000\rangle$. This method, applied to all words
in eq. (\ref{1a}), already yields a possible code. However, it is
\begin{figure}[hbt]
\newcounter{cms}
\setlength{\unitlength}{1.0mm}
\begin{picture}(100,75)
\multiput(11,34)(0,-9){4}{\line(1,0){69}}
\put(11,70){\line(1,0){1}}
\put(16,70){\line(1,0){2}}
\put(22,70){\line(1,0){2}}
\put(28,70){\line(1,0){14}}
\put(46,70){\line(1,0){2}}
\put(52,70){\line(1,0){28}}
\put(11,61){\line(1,0){1}}
\put(16,61){\line(1,0){9}}
\put(27,61){\line(1,0){22}}
\put(51,61){\line(1,0){29}}
\put(11,52){\line(1,0){38}}
\put(51,52){\line(1,0){29}}
\put(11,43){\line(1,0){1}}
\put(16,43){\line(1,0){9}}
\put(27,43){\line(1,0){53}}
\put(14,70){\circle{4}}
\put(14,61){\circle{4}}
\put(14,43){\circle{4}}
\put(14,70){\makebox(0,0)[c]{\bf R}}
\put(14,61){\makebox(0,0)[c]{\bf R}}
\put(14,43){\makebox(0,0)[c]{\bf R}}
\put(14,34){\circle{4}}
\put(14,25){\circle{4}}
\put(14,16){\circle{4}}
\put(14,7){\circle{4}}
\put(14,32){\line(0,1){4}}
\put(14,23){\line(0,1){4}}
\put(14,14){\line(0,1){4}}
\put(14,5){\line(0,1){4}}
\put(20,70){\circle{4}}
\put(20,70){\makebox(0,0)[c]{$\pi$}}
\put(20,61){\circle*{2}}
\put(20,52){\circle*{2}}
\put(20,43){\circle*{2}}
\put(20,43.5){\line(0,1){8}}
\put(20,52.5){\line(0,1){8}}
\put(20,61.5){\line(0,1){6.5}}
\put(26,70){\circle{4}}
\put(26,70){\makebox(0,0)[c]{$\pi$}}
\put(26,61){\circle{2}}
\put(26,52){\circle*{2}}
\put(26,43){\circle{2}}
\put(26,44){\line(0,1){7}}
\put(26,53){\line(0,1){7}}
\put(26,62){\line(0,1){6}}
\put(32,70){\circle*{2}}
\put(32,52){\circle{4}}
\put(32,54){\line(0,1){16.5}}
\put(32,50){\line(0,1){4}}
\put(38,52){\circle{4}}
\put(38,43){\circle*{2}}
\put(38,43.5){\line(0,1){6.5}}
\put(38,50){\line(0,1){4}}
\put(44,70){\circle{4}}
\put(44,70){\makebox(0,0)[c]{$\pi$}}
\put(44,61){\circle*{2}}
\put(44,52){\circle*{2}}
\put(44,43){\circle*{2}}
\put(44,43.5){\line(0,1){8}}
\put(44,52.5){\line(0,1){8}}
\put(44,61.5){\line(0,1){6.5}}
\put(50,70){\circle{4}}
\put(50,70){\makebox(0,0)[c]{$\pi$}}
\put(50,61){\circle{2}}
\put(50,52){\circle{2}}
\put(50,43){\circle*{2}}
\put(50,44){\line(0,1){7}}
\put(50,53){\line(0,1){7}}
\put(50,62){\line(0,1){6}}
\put(56,70){\circle*{2}}
\put(56,7){\circle{4}}
\put(56,9){\line(0,1){60.5}}
\put(56,5){\line(0,1){4}}
\put(62,61){\circle*{2}}
\put(62,16){\circle{4}}
\put(62,18){\line(0,1){42.5}}
\put(62,14){\line(0,1){4}}
\put(68,52){\circle*{2}}
\put(68,25){\circle{4}}
\put(68,27){\line(0,1){24.5}}
\put(68,23){\line(0,1){4}}
\put(74,43){\circle*{2}}
\put(74,34){\circle{4}}
\put(74,36){\line(0,1){6.5}}
\put(74,32){\line(0,1){4}}
\put(5,70){\makebox(0,0)[c]{$|0\rangle$}}
\put(5,61){\makebox(0,0)[c]{$|0\rangle$}}
\put(5,52){\makebox(0,0)[c]{$|\psi\rangle$}}
\put(5,43){\makebox(0,0)[c]{$|0\rangle$}}
\put(5,34){\makebox(0,0)[c]{$|0\rangle$}}
\put(5,25){\makebox(0,0)[c]{$|0\rangle$}}
\put(5,16){\makebox(0,0)[c]{$|0\rangle$}}
\put(5,7){\makebox(0,0)[c]{$|0\rangle$}}
\end{picture}\\
Figure 1: The encoding network: {\bf R} describes a one bit rotation 
which takes
$|0\rangle \rightarrow (|0\rangle + |1\rangle)/\sqrt{2}$ and 
$|1\rangle \rightarrow (|0\rangle - |1\rangle)/\sqrt{2}$. An encircled 
cross denotes a NOT operation while a dot denotes a control bit. For a 
filled circle the
operation is carried out if the control bit is $1$; for an empty circle 
the operation is carried out if the control bit is $0$. A circle with a 
$\pi$ represents multiplication with phase $exp(i\pi)$. Qubit 3 is in 
the state $|\psi\rangle$ that we wish to encode, while all other qubits 
are initially in their ground state $|0\rangle$.
\end{figure}
possible 
to shorten the code by removing bits $5$ and $6$ from every word. This 
then yields eq. (\ref{8}) and analogously eq. (\ref{9}).
Subsequently a computer search was made for
potentially shorter codes; this revealed no such codes, so we conclude 
that $n=8$ qubits is the minimum number required for the task of correcting 
one general error while errors due to the conditional time evolution are
corrected perfectly. In the following we present some interesting properties
of the code and demonstrate that it indeed has the claimed error
correction properties. However, this code differs 
in many ways from previously proposed codes. First of all, it violates the
conditions given for quantum error correcting codes in \cite{Ekert1}
thereby showing that these conditions are overly restrictive. As these
conditions were used to derive the inequality eq. (\ref{3}), their violation
indicates that there might exist codes that require less qubits than 
expected from eq. (\ref{3}). However, we did not yet succeed to construct
a code that violates eq. (\ref{3}). One should also realize that the codewords 
in the code eqs. (\ref{8})-(\ref{9}) do {\em not} form a linear code as this 
would imply that $|00000000\rangle$ is a codeword which in turn would render 
impossible the task of constructing a code with codewords of equal excitation.
Nevertheless, the codewords of $|0_L\rangle$ form a coset of a linear 
code. The coset leader is $|00001111\rangle$. This contrasts slightly 
with other codes such as those presented in
\cite{Shor2,Steane1,Steane2,Laflamme1}. The codewords of the code
(\ref{1a})-(\ref{1b}) for example form a linear code. Given the initial state
$|\psi\rangle=\alpha |0\rangle + \beta |1\rangle$, we obtain the  
code eqs. (\ref{8})-(\ref{9}) using the network given in fig. 1.
To correct the error that may have appeared we first apply the encoder 
in the reverse direction (right to left). After the application of the 
decoder, the third qubit 
contains information about the encoded state while the remaining $7$ 
qubits contain the error syndrome, from which one can infer the type 
and location of the error. We measure the qubits of the error syndrome 
and apply, according to the result of our measurement, a suitable 
unitary operation on qubit $3$. We assume that after the measurement 
all the other qubits are reset to their ground state $|0\rangle$ so 
that, in principle, we can re-encode the state again using the same 
qubits.
\begin{center}
\renewcommand{\arraystretch}{.7} 
\begin{tabular}{|c||c|c|}\hline
Error         & $\;\;$Error syndrome$\;\;$ &          
State of qubit 3           \\ \hline\hline
 None         & $0000000$      
& $\;\;\,\alpha|0\rangle + \beta |1\rangle$ \\ \hline\hline
$P_1$         & $1000000$      
& $\;\;\,\alpha|0\rangle + \beta |1\rangle$ \\ \hline
$P_2$         & $0100000$      
& $\;\;\,\alpha|0\rangle + \beta |1\rangle$ \\ \hline
$P_4$         & $0010000$      
& $\;\;\,\alpha|0\rangle + \beta |1\rangle$ \\ \hline
$A_5$         & $0001000$      
& $\;\;\,\alpha|0\rangle + \beta |1\rangle$ \\ \hline
$A_6$         & $0000100$      
& $\;\;\,\alpha|0\rangle + \beta |1\rangle$ \\ \hline
$A_7$         & $0000010$      
& $\;\;\,\alpha|0\rangle + \beta |1\rangle$ \\ \hline
$A_8$         & $0000001$      
& $\;\;\,\alpha|0\rangle + \beta |1\rangle$ \\ \hline \hline
$P_3$         & $1010000$      
& $\;\;\,\alpha|0\rangle - \beta |1\rangle$ \\ \hline
$A_2$         & $0010010$      
& $\;\;\,\alpha|0\rangle - \beta |1\rangle$ \\ \hline \hline
$P_6$         & $1010000$      
& $-\alpha|0\rangle + \beta |1\rangle$      \\ \hline
$\;\;A_2P_2\;\;$      & $0110010$      
&  $-\alpha|0\rangle + \beta |1\rangle$ \\ \hline
$A_6P_6$      & $1010100$      
& $-\alpha|0\rangle + \beta |1\rangle$ \\ \hline\hline
$P_5$         & $0010000$      
& $-\alpha|0\rangle - \beta |1\rangle$ \\ \hline
$P_7$         & $0100000$     
 & $-\alpha|0\rangle - \beta |1\rangle$ \\ \hline
$P_8$         & $1000000$      
& $-\alpha|0\rangle - \beta |1\rangle$ \\ \hline
$A_5P_5$      & $0011000$      
& $-\alpha|0\rangle - \beta |1\rangle$ \\ \hline
$A_7P_7$      & $0100010$      
& $-\alpha|0\rangle - \beta |1\rangle$ \\ \hline
$A_8P_8$      & $1000001$      
& $-\alpha|0\rangle - \beta |1\rangle$ \\ \hline\hline
$A_1P_1$      & $1110001$      
& $\;\;\,\beta|0\rangle + \alpha|1\rangle$  \\ \hline
$A_4P_4$      & $1011000$      
& $\;\;\,\beta|0\rangle + \alpha|1\rangle$  \\ \hline\hline
$A_3P_3$      & $1110100$      
& $\;\;\,\beta|0\rangle - \alpha|1\rangle$  \\ \hline\hline
$A_1$         & $0110001$     
& $-\beta|0\rangle - \alpha|1\rangle$ \\ \hline
$A_3$         & $0100100$      
& $-\beta|0\rangle - \alpha|1\rangle$ \\ \hline
$A_4$         & $1001000$      
& $-\beta|0\rangle - \alpha|1\rangle$ \\ \hline
\end{tabular}
\end{center}
Table 1: 
One obtains an error syndrome, ie. the state of all qubits except 
qubit $3$, depending on the error that occurred and the place in which it 
occurred. $P_i$ indicates a sign change of the upper level of qubit $i$, 
$A_i$ an amplitude error which is given by the transformation 
$|0\rangle \leftrightarrow |1\rangle$. The product of both
applied to the same qubit gives the third kind of error. Note that the
error syndrome is not able to distinguish between $P_i$ and $P_{9-i}$ 
which leads to global phases in some of the corrected states.
This table does not take into account that before and after the error a 
conditional time evolution takes place.\\[.25cm]
In table 1 we give all possible outcomes of the measurement 
and the corresponding  state of the third qubit. The necessary unitary 
transformation that has to be applied onto the third qubit is then 
obvious. Careful inspection of table 1 reveals that this error 
correction scheme has, for some errors, a slightly different effect 
than expected. Take for example a phase errors $P_1$ on bit $1$ and compare 
with the effect of a phase error $P_8$ on bit $8$. We observe that they 
both lead to the same error syndrome but that the resulting state 
differs by a global phase $-1$. Therefore it is not possible to correct 
both states in such a way that they go over to the initial state. After 
the correction they differ by a global phase $-1$. But this also
shows that the dimension of the space ${\cal H}_{code}$ spanned by
the code together with all states that result from it by single errors
is $2\times 21$ and not as expected from eq. (\ref{3}) $2\times 25$.
The latter number results from the considerations of Ekert and 
Macchiavello \cite{Ekert1} who have presented a set of conditions
that have to be satisfied by any quantum error correction code. 
The violation of these conditions by the code eq. (\ref{8})-(\ref{9})
leads to these different predictions for the dimension of ${\cal H}_{code}$.
More general conditions can be derived and it can be checked easily 
that our code satisfies these conditions \cite{ Knill1,Vedral1} while
it violates the conditions given in \cite{Ekert1}.
 
So far we have shown that our code can indeed correct a general single 
error without taking into account the conditional time evolution due 
to spontaneous emission. Now we show that our code is able to correct 
errors due to the conditional time evolution perfectly, ie. to all orders.
For our code given in eqs. 
(\ref{8})-(\ref{9}) the conditional time evolution under the assumption 
that no spontaneous emission has taken place is generated by the 
effective Hamilton operator
\begin{equation}
	H_{eff} = \sum_{i=1}^{8} -i\hbar\Gamma \sigma_{11}^{(i)} \;\; .
	\label{10}
\end{equation}
If the code undergoes a conditional time evolution before it experiences 
an error like e.g. a spontaneous emission, it is obvious 
that the code eqs. (\ref{8})-(\ref{9}) will work properly, as it is 
invariant under the conditional 
time evolution $exp(-i H_{eff} t/\hbar)$. However, it is not so obvious 
that the code corrects general single errors that occur {\em before}
or in between the conditional time evolution. As we do not know the 
time at which the general error occurs, this situation will almost 
certainly occur and has to be examined. If the error was a phase error, 
then no problem will occur, as this error does not change the excitation 
of the state. However, for amplitude errors or a combination of amplitude 
and phase errors we have to investigate the code more closely. The 
problem is that, for example after an amplitude error in the first qubit, 
we obtain 
\begin{eqnarray}
	A_1|0_L\rangle &=& \nonumber\\
	&&\hspace*{-1.5cm}  |10001111\rangle - |11110001\rangle + 
			    |11001101\rangle + |10110011\rangle
	\nonumber\\
	         &&\hspace*{-1.5cm} +|0110100\rangle - 
                           |00010110\rangle + |01010100\rangle
			  +|00101010\rangle\;\; .
	\label{11}
\end{eqnarray}
Now the code words have a different degree of excitation so that their 
relative weights will change during the subsequent conditional time 
evolution. However, for $|\psi_L\rangle=\alpha|0_L\rangle + \beta|1_L\rangle$ 
we have the relations
$|\psi_L\rangle=\alpha|0_L\rangle + \beta|1_L\rangle$
\begin{eqnarray}
	e^{-i H_{eff} t/\hbar} A_i |\psi_L\rangle &=& \nonumber\\
	&&\hspace*{-2.5cm}
	\frac{1}{2} e^{-3\Gamma t} \left\{ (1 + e^{-2\Gamma t}) A_i 
	- (1 - e^{-2\Gamma t}) A_i P_i \right\} |\psi_L\rangle
	\label{12}
\end{eqnarray}
and
\begin{eqnarray}
	e^{-i H_{eff} t/\hbar} A_i P_i |\psi_L\rangle &=& \nonumber\\
	&&\hspace*{-3.cm}
	\frac{1}{2} e^{-3\Gamma t} \left\{ -(1 - e^{-2\Gamma t}) A_i 
	+ (1 + e^{-2\Gamma t}) A_i P_i \right\} |\psi_L\rangle \;\; .
	\label{13}
\end{eqnarray}
Eq. (\ref{12}) shows that after an amplitude error $A_i$ on the $i$th 
qubit, the conditional time evolution transforms the state into a 
superposition of a state without conditional time evolution after 
this amplitude error, and a state without conditional time evolution 
obtained after a combined amplitude and phase error $A_iP_i$ on the 
$i$th qubit. Inspecting table 1 we see that both errors $A_i$ and 
$A_iP_i$ lead to a different error syndrome. A measurement of the syndrome 
will then indicate one or the other error, $A_i$ or $A_iP_i$, which 
can then be corrected. Therefore the code (\ref{8})-(\ref{9}) corrects 
properly even if the error is followed by a conditional time evolution.
\section{Conclusions} 
We conclude that the code presented here is able to correct 
a single general error {\em and} in addition errors due to the 
conditional time evolution to {\em arbitrary} order. It is the 
first code proposed so far that can correct a general errors to first order
{\em and} a special kind of errors to all orders. This is an 
interesting result as it shows that it is possible to correct 
special kinds of errors to all orders. As some errors are more 
frequent than others it would be in our interest to correct 
those errors to higher order than less frequently occurring 
errors. We have adapted our code to errors due to the conditional 
time evolution between spontaneous emissions. Other applications 
will require different adaptions. The code presented here (similar to the
one given in \cite{Shor2}) violates 
the conditions for quantum codes given in \cite{Ekert1} which shows 
that these conditions are overly restrictive, as they exclude codes 
like the one presented here
that map different errors onto the same error syndromes. This can 
lead to the construction of shorter quantum error correction codes 
than expected from the quantum sphere packing bound as derived in
\cite{Ekert1}. These results may become important in 
different fields such as quantum computation, the distribution of 
entangled particles and in quantum cryptography 
\cite{Bennett2,Ekert2,Hughes1,Phoenix1}.\\[.75cm]

\noindent
{\bf Acknowledgements}\\[.25cm]
The authors thank A. Ekert, C. Macchiavello and A.M. Steane 
for discussions about quantum error correction. This work was
supported by the European Community, 
the UK Engineering and Physical Sciences Research Council and by a 
Feodor-Lynen grant of the Alexander von Humboldt foundation.


\begin{references}
\bibitem{Shor1} P.W. Shor. {\em In Proc. $35th$ Annual Symposium
on Foundations of Computer Science}, ed. S. Goldwasser. (IEEE 
Computer Society Press, Nov. 1994) pp. 124-134.
%
\bibitem{Cirac1} J.I. Cirac and P. Zoller, Phys. Rev. Lett. {\bf 74},
4091 (1995)
%
\bibitem{Plenio1} M.B. Plenio and P.L. Knight, Phys. Rev. A {\bf 53},
2986 (1996)
%
\bibitem{Plenio2} M.B. Plenio and P.L. Knight, {Proceedings of the 
2nd International Symposium on Fundamental Problems in Quantum Physics}, 1996 
edited by M. Ferrero and A. Van der Merwe (Kluwer, Dordrecht)
%
\bibitem{Shor2} P.W. Shor, Phys. Rev. A {\bf 52}, R2493 (1995)
%
\bibitem{Chuang1} I.L. Chuang and R. Laflamme, {\em Quantum Error 
Correction by Coding.}, preprint quant-ph/9511003
%
\bibitem{Calderbank1} A.R. Calderbank and P.W. Shor, {\em Good Quantum 
Error-Correcting Codes Exist.} preprint quant-ph/9512032
%
\bibitem{Steane1} A.M. Steane, Phys. Rev. Lett. {\bf 77}, 793 (1996) 
%
\bibitem{Steane2} A.M. Steane, {\em Multiple Particle Interference and
Quantum Error Correction.} lanl e-print quant-ph/9601029
%
\bibitem{Laflamme1} R. Laflamme, C. Miquel, J.P. Paz, and W.H. Zurek,
{\em Perfect Quantum Error Correction Code.} preprint quant-ph/9602019
%
\bibitem{Ekert1} A. Ekert and C. Macchiavello, {\em Error Correction in 
Quantum Communication.} preprint quant-ph/9602022, Oxford University
%
\bibitem{Vaidman1} L. Vaidman, L. Goldenberg, and S. Wiesner, {\em
Error Prevention Scheme with Four Particles},
lanl e-print quant-ph/9603031
%
\bibitem{ Knill1} E. Knill and R. Laflamme, {\em A theory of quantum 
error-correcting code}, lanl e-print quant-ph/9604015
%
\bibitem{Bennett1} C.H. Bennett, D.P. DiVincenzo, J.A. Smolin, and
W.K. Wootters, {\em Mixed state entanglement and quantum error-correction
codes}, lanl e-print quant-ph/9604024
%
\bibitem{Barenco1} A. Barenco, A. Berthiaume, D. Deutsch, A. Ekert, R. Josza,
and C. Macchiavello, {\em Stabilisation of quantum computations by symmetrisation}, lanl e-print quant-ph/9604028
%
\bibitem{Calderbank2} A.R. Calderbank, E.M. Rains, P.W. Shor, and N.J.A. Sloane, lanl e-print quant-ph/9605005
%
\bibitem{Shor3} P.W. Shor, {\em Fault-Tolerant Quantum Computation}, 
lanl e-print quant-ph/9605011
%
\bibitem{Steane3} A.M. Steane, {\em Simple Quantum Error Correcting Codes}, 
lanl e-print quant-ph/9605021
%
\bibitem{DiVincenzo1} D.P. DiVincenzo and P.W. Shor, {\em Fault-Tolerant Error
Correction with Efficient Quantum Codes}, lanl e-print quant-ph/9605031
%
\bibitem{Gottesmann2} D. Gottesmann, {\em Pasting Quantum Codes}, lanl 
e-print quant-ph/9607027
%
\bibitem{ Gottesmann1} D. Gottesmann, Phys. Rev. A {\bf 54}, September 1996
%
\bibitem{Cleve1} D. Gottesmann, {\em Efficient Computations of Encodings for
Quantum Error Correction}, lanl e-print quant-ph/9607030
%
\bibitem{Dalibard1} J. Dalibard, Y. Castin, and K. M{\o}lmer, Phys. Rev.
Lett. {\bf 68}, 580 (1992)\\
G. C. Hegerfeldt and T. S. Wilser,
{\em Proceedings of the II. International Wigner
Symposium, Goslar 1991}, H.D. Doebner, W. Scherer, and F. Schroeck,
Eds., World Scientific, Singapore 1992\\
H.J. Carmichael, {An Open Systems Approach to Quantum Optics.} Lecture 
Notes In Physics, (Springer, Berlin 1993)
%
\bibitem{Pless1} V. Pless, {\em Introduction to the Theory of Error-Correcting
Codes}, John Wiley \& Sons 1982
%
\bibitem{Mabuchi1} H. Mabuchi and P. Zoller, Phys. Rev. Lett. {\bf 76}, 3108 (1996)
%
\bibitem{Vedral1} V. Vedral, M. Rippin, and M.B. Plenio, submitted to 
Phys. Rev. A 1996
%
\bibitem{Bennett2} C.H. Bennett and G. Brassard, in {\em Proceedings
if IEEE Conference on Computers, Systems and Signal Processing}, p.175
(IEEE 1984)
%
\bibitem{Ekert2} A. Ekert, Phys. Rev. Lett. {\bf 67}, 661 (1991)
%
\bibitem{Hughes1} R.J. Hughes, D.M. Alde, P. Dyer, G.G. Luther, G.L. Morgan
and M. Schauer, Cont. Physics {\bf 36}, 149 (1995)
%
\bibitem{Phoenix1} S.J.D. Phoenix and P.D. Townsend, Cont. Physics 
{\bf 36}, 165 (1995)
%
\end{references}
\end{document}